\documentclass[twocolumn,floatfix]{revtex4}

	\usepackage{amssymb}
	\usepackage{amsmath}
	\usepackage{epsfig}
	\usepackage{graphicx}

\begin{document}
\title{Scale Invariance in Global Terrorism}
\author{Aaron Clauset and Maxwell Young}
\affiliation{Department of Computer Science, \\ University of New Mexico, Albuquerque NM 87131 \\
{\tt (aaron,young)@cs.unm.edu} }
\date{\today}

\begin{abstract}
Traditional analyses of international terrorism have not sought to explain the emergence of rare but extremely severe events. Using the tools of extremal statistics to analyze the set of terrorist attacks worldwide between 1968 and 2004, as compiled by the National Memorial Institute for the Prevention of Terrorism (MIPT), we find that the relationship between the frequency and severity of terrorist attacks exhibits the ``scale-free'' property with an exponent of close to two. This property is robust, even when we restrict our analysis to events from a single type of weapon or events within major industrialized nations. We also find that the distribution of event sizes has changed very little over the past 37 years, suggesting that scale invariance is an inherent feature of global terrorism.
\end{abstract}
\maketitle

Although terrorism has a long historical relationship with politics~\cite{Congleton01}, only in the modern era have small groups of non-state actors had access to extremely destructive weapons~\cite{FBI99, Shubik97}, particularly chemical or explosive agents. This dramatic increase in destructive power has allowed severe terrorist attacks such as the \mbox{March 20 1995} release of the Sarin nerve agent in a Tokyo subway which injured or killed over $5000$, the \mbox{August 7 1998} car bombing in Nairobi, Kenya which injured or killed over $5200$, or the more well known attack on \mbox{September 11 2001} in New York City which killed $2823~$\footnote{National Memorial Institute for the Prevention of Terrorism (MIPT) Terrorism Knowledge Base (Jan. 2005). {\tt www.tkb.org }}. Typical analyses of patterns of terrorist events have treated such rare but severe attacks as outliers, and generally focused attention only on the group of subjectively defined ``significant'' events~\cite{FBI99,State03}. We show here, by examining data over the past 37 years, that discounting extremal events as special cases ignores a significant pattern in terrorism.

Using the tools of extremal statistics, we characterize the relationship between the {\em severity} and frequency of terrorist events. By severity, we simply mean the number of individuals injured or killed by an attack. We show that this relationship may be well-characterized by the simple mathematical function, the power law $P(x)\sim x^{-\alpha}$, where $\alpha$ is the scaling exponent. Such a pattern is said to be ``scale invariant,'' and is ubiquitous in nature, appearing in the frequency of word usage in a variety of languages (known more commonly as Zipf's Law), the number of citations per scientific paper, the population of cities, the magnitude of earthquakes, the net worth in US dollars of individuals, etc. (see, for example,~\cite{Newman04}). For this work, the most relevant power law is the well established one governing the relationship between the frequency and intensity~\footnote{Intensity has several definitions, the simplest being casualties per $10~000$ people in the warring nations.} of wars~\cite{Richardson48, Richardson60, Levy83, Roberts98}, to which we will return in a later section.

Although many organizations track such attacks worldwide, few provide their data publicly or in anything but an aggregate form. The MIPT database appears to be unique in its comprehensive detail, containing, as of January 2005, records of over $19~907$ terrorist events in 187 countries worldwide between 1968 and 2004. Of these, $7~088$ resulted in at least one person being injured or killed. The MIPT database is itself the compilation of the RAND Terrorism Chronology 1968-1997, the RAND-MIPT Terrorism Incident database (1998-Present), the Terrorism Indictment database (University of Arkansas \& University of Oklahoma), and DFI International's research on terrorist organizations. Each record includes the date, target, city (if applicable), country, type of weapon used, terrorist group responsible (if known), number of deaths (if known), number of injuries (if known), a brief description of the attack and the source of the information. 

\vspace{3mm}\noindent {\bf \large Global Patterns}

\begin{figure*}[t] 
\begin{center}
\begin{tabular*}{17.5cm}{lll}
\includegraphics[scale=0.318]{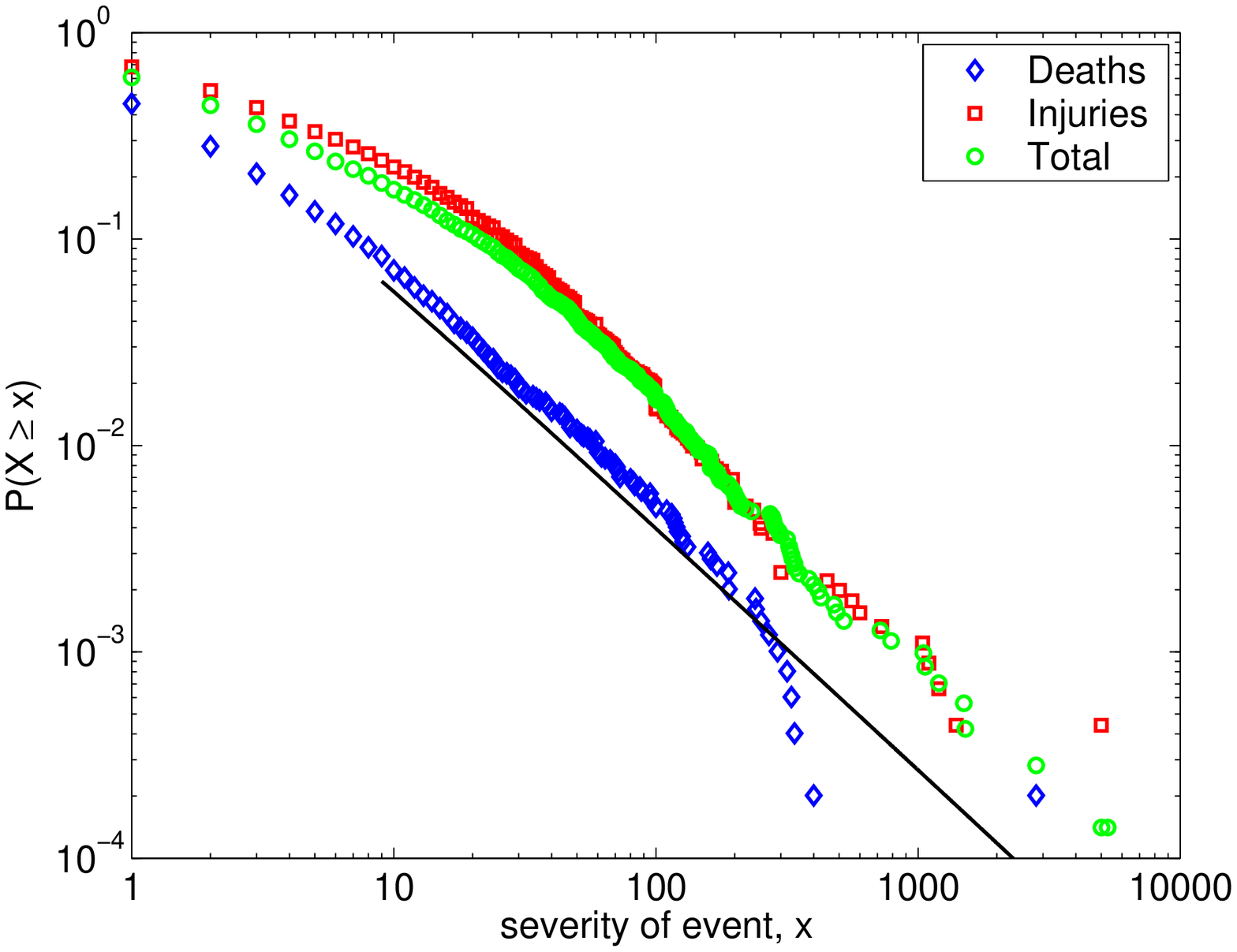} &
\includegraphics[scale=0.318]{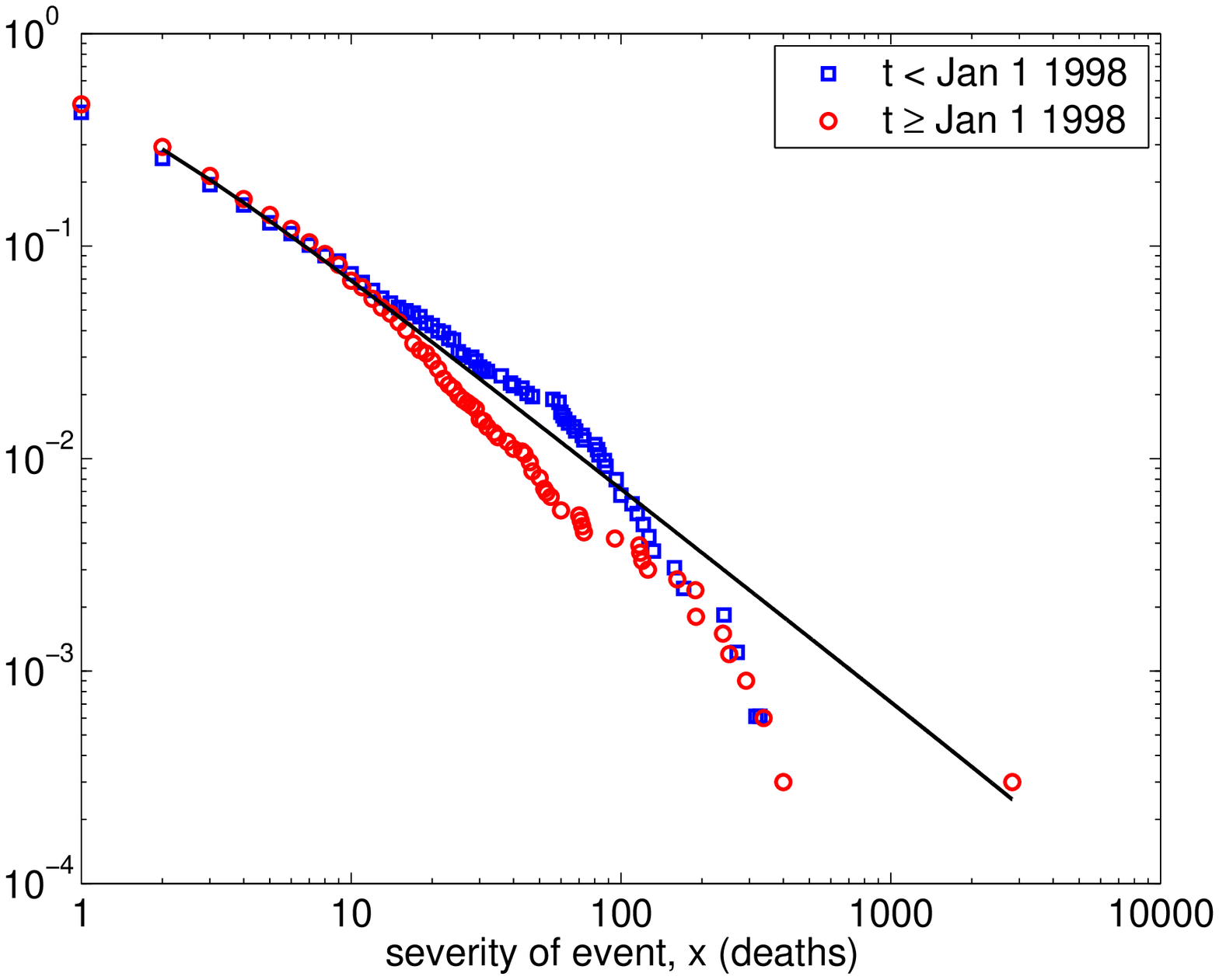} &
\includegraphics[scale=0.318]{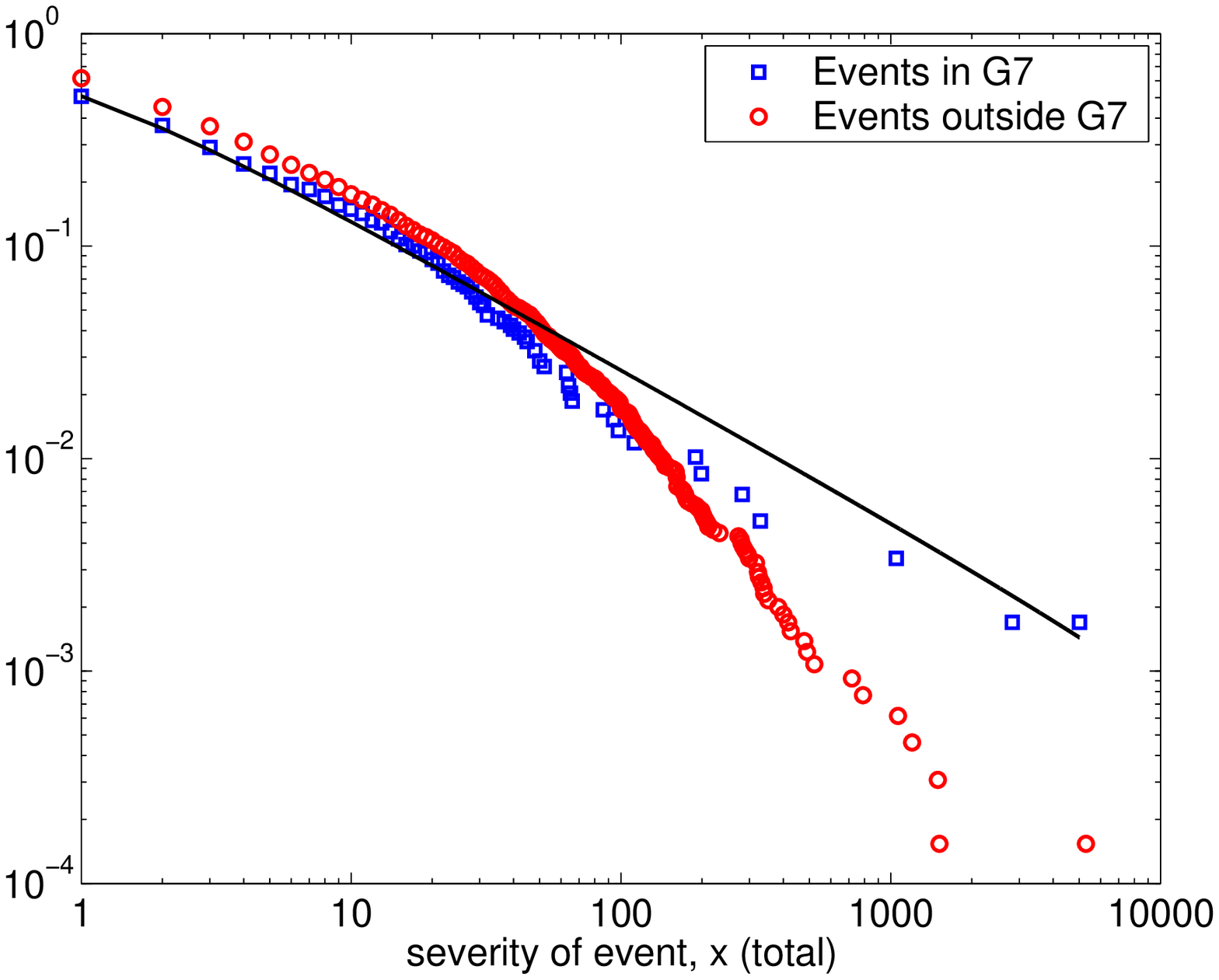} \\
\end{tabular*}
\end{center}
\caption{The distributions $P(X\geq x)$ of attack severity for attacks worldwide between 1968 and 2004 (a) by injuries, deaths and their aggregation, (b) for before and after the management of the database was assumed by MIPT, and (c) for events inside versus outside the G7 industrialized nations. Solid lines are guides indicating the maximum likelihood power laws given in Table~\ref{table:summary}.}
\label{fig:terrorism}
\end{figure*}

\noindent Tabulating the event data as a histogram of severity (injuries, deaths and their aggregation, greater than zero), we show the cumulative distribution functions \mbox{$P(X\geq x)$} on log-log axes in Figure~\ref{fig:terrorism}a. The regularity of the scaling in the tails of these distributions suggests that the extremal events are not outliers, but are instead in concordance with a global pattern in terrorist attacks. This scaling exists in spite of strong heterogeneity in the type of weapon, the perpetrating organization, geographic location, political motivation behind the attack, etc. 

Assuming for the moment that the events are i.i.d., we hypothesize that the distributions follow power laws of the form $P(x)\sim x^{-\alpha}$ above some minimum value $x_{min}$. We use the log-likelihood function for the discrete power law with minimum value $x_{min}$,
\begin{align*}
\mathcal{L} &= \ln P(x|\alpha) \\ 
 & = - \sum_{i=x_{min}}^{n} \left( \alpha \ln x_{i} + \ln\!\left[\zeta(\alpha) - \sum_{j=1}^{x_{min}-1}x_{j}^{-\alpha} \right] \right)
 \end{align*}
to fit the distributions, where $\zeta(\alpha)$ is the Reimann zeta function. Our procedure is thus: we bootstrap a numeric maximization of $\mathcal{L}$ to estimate the scaling parameter $\alpha$, and minimize the D-statistic of the Kolmogorov-Smirnov goodness-of-fit test to select the parameter $x_{min}$. Using Monto Carlo methods to generate a table of p-values for the Kolmogorov-Smirnov statistical test for power law with the estimated $\alpha$ and $x_{min}$, we find that there is insufficient evidence to reject the power law as a model of the tails of these distributions. We will see in a later section, however, that these distributions are the composition of several distinct power laws with different scaling parameters and ranges, which, in turn, causes some roughness in the power law model. Although we do find sufficient evidence to reject the log-normal distribution~\cite{Serfling02} hypothesis in all cases ($p_{KS}<0.05$), we cannot completely rule out Type I/II errors as asymptotic scaling tests are quite sensitive to the range and number of observations. The consideration of other heavy-tailed distributions, e.g., the \mbox{q-exponential} $e_{q}^{-\alpha x}$ or the stretched exponential $e^{-\alpha x^{\beta}}$, which may result in a better fit of the lower tail of the distributions, is beyond the scope of this work. Table~\ref{table:summary} summarizes the statistics and power law models for all distributions shown in Figure~\ref{fig:terrorism}.

A few brief comments are in order. The form of our power law model ignores all data below $x_{min}$; thus, we only model the upper tails and say nothing about the shallow scaling off the lower tails (e.g., injuries, total). Additionally, within the range defined by $x_{min}$, the power law is not as clean as, for example, that of earthquakes~\cite{Newman04}. In a system as complex as global terrorism, such irregularities are to be expected; however, the appearance of scale invariance is not. Finally, Figure~\ref{fig:terrorism}b and c show two additional views of global terrorism, which we will explain in the subsequent section. 

\vspace{3mm}\noindent {\bf \large Bias, Trends and Components}

\noindent A significant event during the 37 years over which the MIPT database spans was the assumption of data management by the MIPT from the RAND corporation in 1998. This event raises the natural question of whether any fundamental bias has been introduced into the data as a result of differing management practices or changes to the criteria used to judge the admissibility of an event, differences which may result in the regular scaling observed in Figure~\ref{fig:terrorism}a. In this section, we explore this question and characterize the evolution of the distribution as a function of time.

\begin{table}[t]
\begin{center}
\begin{tabular}{|l|cccccc|cccccc|}
\hline
Distribution & $\langle x\rangle$ & & std. & & $x_{max}$ & & & $\alpha$  & & $x_{min}$ & & $p_{KS}\geq$ \\
\hline
Injuries  & 14.60 & & 114.82 & & 5000 & & & 2.40(6) & & 34 & & 0.658 \\
\hline
Deaths   & 5.13 & & 43.37 & & 2823 & & & 2.21(6)  & & 8 & & 0.632 \\
$t<1998$    &  5.18 & & 19.21 & & 329 & & & 1.92(3)  & & 1 & & 0.999 \\
$t\geq1998$  & 5.11 & & 51.20 & & 2823 & & & 2.00(2)  & & 2 & & 0.999 \\
\hline
Total       &  12.70 & & 103.38 & & 5291 & & & 2.17(4) & & 26 & & 0.404 \\
G7           &  22.66 & & 241.18 & & 5012 & & & 1.71(3)  & & 1 & & 0.997 \\
Non-G7  &  11.80 & & 79.85 & & 5291 & & & 2.5(1)  & & 86 & & 0.971 \\
\hline
\end{tabular}
\end{center}\caption{A summary of the distributions shown in Figure~\ref{fig:terrorism}, and their maximum likelihood power laws. The value in parentheses is the estimated error in the last digit of the scaling exponent. }
\label{table:summary}
\end{table}%

\noindent {\bf Reporting Bias.} In Figure~\ref{fig:terrorism}b, we plot the severity distributions (deaths) for all events before ($2~304$, or $33\%$) and after ($4~784$, or $67\%$) the change in management, which we take to occur on \mbox{1 January 1998}. Notably, the distributions are very similar and the scaling robust to the significant increase in the frequency of events after 1998. We find that both distributions follow a power law with exponent $\alpha\approx2$ and that, in spite of differences in the mid- and upper-tail, the slope changed by only $4\%$. Although we don't model it as such, the earlier distribution may be best fit by a power law with exponential cutoff beginning at $x=87$; such a cutoff could be the result of technological constraints. Generally, the appearance of the scale invariance itself seems unlikely to be the result of changes in database management. Unfortunately, we have no way of accounting for any human-bias in the decision of an event's admissibility. However, an analysis of the smaller but independent database maintained by the International Policy Institute for Counter-Terrorism's (ICT)~\footnote{B. Ganor et al., International Policy Institute for Counter-Terrorism (Jan. 2005). {\tt www.ict.org.il }}, which contains 1417 events between May 1980 and December 2002, yields similar results (not shown), suggesting that the scaling in the upper tails is a robust feature of global terrorism.

\begin{figure} [t]
\begin{center}
\includegraphics[scale=0.45]{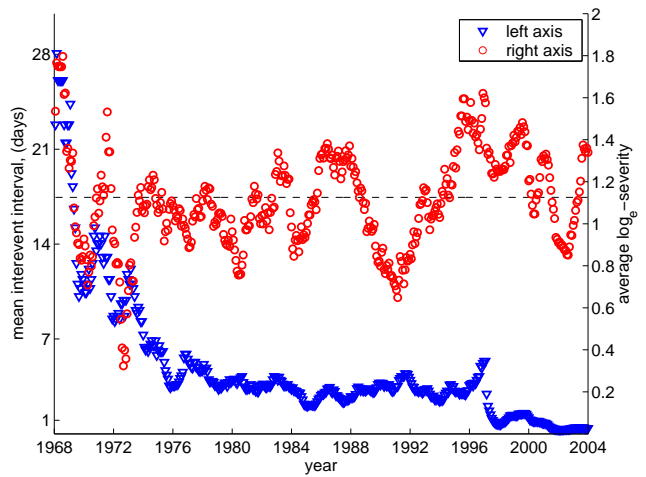}
\caption{Time series showing (left axis) the average inter-event interval over a sliding window of 12 months, and (right axis) the average logarithm of the total severity of attacks over the same window of 12 months. The dashed line shows the average log-severity for all events. Over the course of the 37 years, we find a slight, but statistically significant, linear trend in the breadth of $p(x)$, with slope $m=4.2\times 10^{-4}$, despite large changes in the frequency of events.  }
\label{fig:timeseries}
\end{center}
\end{figure}

\noindent {\bf Inter-event Interval.} We now consider the evolution of the severity distribution over time. As mentioned above, the majority of events which killed or injured at least one person have occurred since $1998$. Figure~\ref{fig:timeseries} (left axis) shows the mean time (in days) between recorded events, taken over a 12 month sliding-window.
The decrease in the mean inter-event interval is striking, falling from approximately 28 days in early 1968 to less than 12 hours in 2004. The precipitous drop in 1998 is especially notable, and suggests that when the database changed hands, the admission criteria may have become more permissive or a significantly larger number of events were being evaluated. However, Figure~\ref{fig:timeseries} illustrates that the decrease in inter-event time has been a largely continuous trend over the entire lifetime of the database. We note that this increase in the frequency of recorded events worldwide is consistent with findings by the United States Department of State~\cite{State03}. Thus, while it seems plausible that the change in maintenance did result in some increase in reporting frequency, we cannot rule out the possibility that there has been a genuine increase in the frequency of attacks in recent years.

\noindent {\bf Global Trends.} Given that the distributions we observe in Figure~\ref{fig:terrorism}a and b are a collection of events over time, we may naturally wonder if each event was drawn from the same distribution (i.e., the severity distribution is stable), or if the distribution has changed in some way. The right-axis in Figure~\ref{fig:timeseries} shows the average log-severity of events within the same sliding window of 12 months, over the 37 years of data. Let us assume that each event was drawn independently from the distribution $p(x)$. We then expect the time series to fluctuate about the average log-severity for the entire time period, which is $\langle\ln(s)\rangle = 1.126$. This appears to be the case, although we also measure a slight linear trend in this function, by least-squares, with slope $m= 4.2\times 10^{-4}$. Using Monte Carlo simulation, we test the likelihood that this linear trend is the result of random fluctuations arising from the variation in sampling frequency. To do this, we simulate a new sequence of events by drawing a new severity value from the hypothesized heavy-tailed distribution $p(x)$ (Table~\ref{table:summary}) for each observed time; we then compute the window-averaged log-severity as above and measure its linear component $m^{*}$. Repeating this process many times, we find that we may reject the random-fluctuation hypothesis ($p_{MC}<0.013$). Bootstrapping the observed distribution yields a similar conclusion. Thus, we may say with some confidence that while the severity distribution has certainly not changed much over the past 37 years, the slight linear trend is unlikely to be the result of random fluctuations due to sampling, or the increased frequency of events in recent years.

\noindent {\bf Component Distributions.} As mentioned above, the richness of the event meta-data allows for many views of global terrorism. Here, we focus specifically on those which are relevant to the appearance of scale invariance in the severity distributions. Dividing events into categories for those occurring within the major industrialized nations, i.e., Canada, France, Germany, Italy, Japan, the United Kingdom and the United States, known collectively as the G7 (590 events over 37 years), and those occurring throughout the rest of the world ($6~498$ events, or $92\%$), we plot the corresponding severity distributions (total) in Figure~\ref{fig:terrorism}c (also summarized in Table~\ref{table:summary}). Most notable is the substantial difference in the scaling in the upper tails: $\alpha_{G7}=1.71\pm0.03$ versus $\alpha_{non-G7}=2.5\pm0.1$; that is, the largest events are significantly more likely to occur within one of the G7 nations than elsewhere in the world. We have no firm explanation for such a distinction, although it may be the result of technological differences. That is, small groups of non-state actors may have access to a greater degree of destructive potential as a result of industrialization.

\begin{figure} [t]
\begin{center}
\includegraphics[scale=0.45]{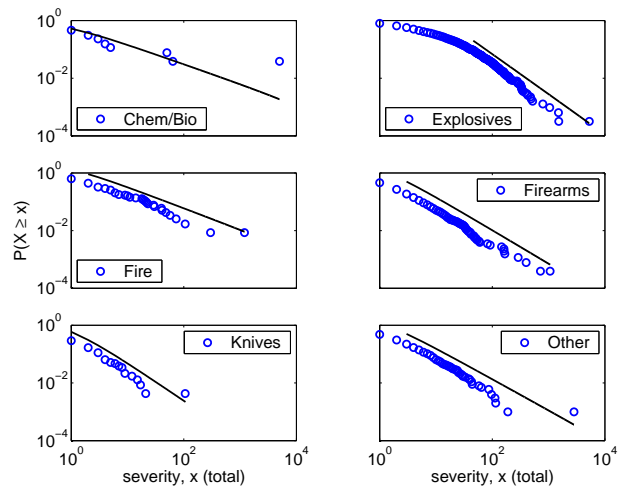}
\caption{Total severity distributions and their corresponding maximum likelihood power laws (see Table~\ref{table:wsummary}) for six weapon types: chemical or biological agents ($0.4\%$ of events), explosives (including remotely detonated devices) ($44.3\%$), fire ($1.7\%$), firearms ($36.2\%$), knives ($3.3\%$) and other (which includes unconventional and unknown weapon types) ($14.2\%$).  }
\label{fig:weapons}
\end{center}
\end{figure}

We may also divide the events into groups based on the type of weapon used; Figure~\ref{fig:weapons} shows the (total severity) distributions of events for chemical or biological weapons, explosives (including remotely detonated devices), fire, firearms, knives and a catch-all category ``other'' (which also includes unconventional and unknown weapons). Surprisingly, these component distributions are all well-modeled by power laws ($p_{KS}\geq0.9$), which we summarize in Table~\ref{table:wsummary}. There are several points to be made by this observation. First, the trend in the lower tails of Figure~\ref{fig:terrorism}a and c (injuries or total, and non-G7, respectively) is now obviously caused by some property of explosives. Additionally, different weapon types clearly exhibit distinct cutoffs, as one might expect, e.g., knives have the smallest associated maximum severity.

\vspace{5mm}\noindent {\bf \large Models and Mechanisms}

\noindent We believe that the scale invariance in global terrorism is related in some way to the power law observed by Richardson in 1948~\cite{Richardson48}, and confirmed independently by~\cite{Levy83,Roberts98}, for the frequency versus intensity of wars. Using a similar maximum likelihood method on the historical war data of Small and Singer~\cite{Small82}, Newman found a scaling parameter of \mbox{$\alpha=1.80\pm0.09$}, which is close to several of those that we measure for global terrorism.

Given the appearance of scale invariance across many views of global terrorism, we may now ask if a simple generative model exists which might explain its origin. We note that the complexity of terrorism makes such a simple explanation unlikely or, at best, highly susceptible to criticism. In this section, we discuss a few power-law mechanisms which are appealing for the system of terrorism/counter-terrorism (see~\cite{Newman04, Mitzenmacher04, Farmer05} for brief surveys of other power-law mechanisms). Although it has been suggested that Richardson's scaling law is the result of a metastability in a geopolitical system~\cite{Cederman03} that has driven itself to a state of self-organized criticality (SOC)~\cite{Bak87}, this hypothesis seems ill-suited to explain the scaling in the severity of terrorist attacks. Another appealing model is that of highly optimized tolerance (HOT)~\cite{Carlson00}; however, because it relies both on risk-neutrality and underlying geometric constraints to produce power laws~\cite{Newman02}, it seems a poor fit for our system. Indeed, the most appealing model, and one that is strongly supported by the data, is one based on the mixing of component distributions such as those in Figure~\ref{fig:terrorism}c and Figure~\ref{fig:weapons}.

In the interest of framing future work in this area, we suggest a short list of criteria by which to judge any proposed general model of global terrorism: a successful model must (in order of importance)

\begin{enumerate}
\item represent an intuitive mechanism by which to generate the size of an actual event (e.g., a competition between states and non-state actors);
\item produce heavy tails with appropriate scaling (Fig.~\ref{fig:terrorism}a);
\item allow for the resulting distributions and dynamics to vary in time; and,
\item account for the differences in scaling caused by technology, e.g., industrial versus non-industrial nations (Fig.~\ref{fig:terrorism}c) and types of weapons (Fig.~\ref{fig:weapons}).
\end{enumerate}

\begin{table}[t]
\begin{center}
\begin{tabular}{|l|cccccc|cccccc|}
\hline
Distribution & $\langle x\rangle$ & & std. & & $x_{max}$ & & & $\alpha$  & & $x_{min}$ & & $p_{KS}\geq$ \\
\hline
Chem/Bio   & 198.73 & & 981.84 & & 5012 & & & 1.8(2) & & 1& & 0.912  \\
Explosives  &  20.42 & & 111.03 & & 5291 & & & 2.38(7)  & & 46 & & 0.894 \\
Fire              &  19.69 & & 113.78 & & 1200 & & & 1.74(9)  & & 2 & & 0.973 \\
Firearms     &  4.29 & & 28.32 & & 1065 & & & 2.17(3)  & & 3 & & 0.997 \\	
Knives        &  2.35 & & 7.28 & & 107 & & & 2.3(1) & & 1 & & 0.999 \\
Other          &  6.85 & & 89.66 & & 2823 & & & 2.07(5)  & & 3 & & 0.991 \\
\hline
\end{tabular}
\end{center}\caption{A summary of the distributions shown in Figure~\ref{fig:weapons}, and their maximum likelihood power laws. The value in parentheses is the estimated error in the last digit of the scaling exponent. }
\label{table:wsummary}
\end{table}%

We will now analyze a highly idealized mathematical model of the competition between non-state actors and states which satisfies these criteria in the most general sense. Let us assume a large population of non-state actors, each of whom is responsible for executing a single event, and let each event's severity be given by a random variable $s$ with distribution $p(s)$, with some maximum value $s_{max}$. Now assume that some, but not all, events are actually executed, perhaps because of collective counter-terrorism actions by states, social factors, random failures, etc. Thus, the distribution of actual events $p(x)$ is given by a sampling of the distribution of potential severities $p(s)$, and may be derived by solving the equation
\[ p(x){\rm d}x = p(s){\rm d}s \enspace . \]
Suppose that, perhaps as a result of factors such as the relative availability of cheap weapons, prevalence of potential targets, technological advances~\cite{Shubik97}, etc., the distribution of potential severities is exponential, $p(s)\sim{\rm e}^{as}$, with \mbox{$a>0$} up to some $s_{max}$. Now suppose that the likelihood of an event being successful is inversely related to its potential severity, so the relationship between $x$ and $s$ might be given by $x\sim {\rm e}^{bs}$ with $b<0$. Under these assumptions, the solution to our general equation above yields a power law $p(x) \sim x^{-\alpha}$, where $\alpha= 1-a/b$. When \mbox{$|a|\approx |b|$}, we derive a power law with exponent \mbox{$\alpha\approx 2$}.

As shown clearly in Figure~\ref{fig:weapons}, the severity distribution for global terrorism is itself composed of several distinct component distributions. Thus, it is appealing to consider mathematical models derived from distribution mixtures, e.g.,
\[ p(x) = \int g(z)\, f_{z}(x)\, {\rm d}z \enspace , \]
where $z$ distinguishes each component distribution $f_{z}(x)$, and $g(z)$ is some mixing function. From this approach, if the $f_{z}(x)$ decay faster than any power law, then $p(x)$ itself will only be a power law when $g(z)$ is a power law~\cite{Sornette00}. Thus, we must still rely upon a power-law generating mechanism, such as the idealized one given above. However, this meta-model has the appealing feature of being able to capture real heterogeneity in the data: if we take $g(z)$ to be some constant, and let $(a/b)_{G7}=1.71$ and $(a/b)_{non-G7}=2.5$, we may recover a distribution which has scaling in the upper tail like that of total severity in Figure~\ref{fig:terrorism}a. An empirical estimate of these ratios would provide a point of validation for our model, but admittedly may be quite difficult to make.

The mixtures model may be even more useful. As we would expect, Figure~\ref{fig:weapons} shows that the severity of an event is governed by the type of weapon used. Thus, the choice of exponential distribution for the potential severity of an event $p(s)\sim {\rm e}^{as}$ must itself be a mixture of weapon-specific distributions. We note also that our simple generative model may be extended to incorporate the slight temporal trend in the average log-severity of events, as illustrated in Figure~\ref{fig:timeseries} by simply letting the parameters $a$ and $b$ vary in time. Finally, it should be noted that our simple model assumes each event is drawn i.i.d. from the underlying distribution; obviously, in the real data, there are likely strong temporal correlations associated with various geopolitical events and policies. A more realistic model might account for these correlations.

\vspace{3mm}
\noindent {\bf \large Conclusions}

\noindent In exploring the distribution of the severity of events in global terrorism, we have found a surprising and robust feature: scale invariance. Traditional analyses of terrorism have typically viewed catastrophic events such as the 1995 truck bombing of the American embassy in Nairobi, Kenya, which killed or injured more than $5~200$, as outliers. However, the property of scale invariance suggests that these are instead a part of a statistically significant global pattern in terrorism. Further, we find little reason to believe that the appearance of power laws in the distribution of the severity of an event is the result of either reporting bias or changes in database management. This suggests that the power law distribution, with $\alpha\approx2$, is an inherent feature of terrorism and counter-terrorism. Indeed, the severity distribution itself has changed very little over the past 37 years of recorded events (Fig.~\ref{fig:timeseries}), in spite of a dramatic increase in the frequency of events. This small growth in the breadth of the severity distribution may be the result of technological changes, such as the power and availability of cheap explosives and firearms.

Surprisingly, the scale-invariance result extends beyond the total collection of events. When we examine the distributions for major industrialized nations versus the rest of the world, we find that heavy tails are present in both (Fig.~\ref{fig:terrorism}c), but with substantially different exponents: $\alpha_{G7}=1.71\pm0.03$ versus $\alpha_{non-G7}=2.5\pm0.1$. That is, while events occur much less frequently in major industrialized nations, when they do, they are much more severe (on average) than events outside those nations. Additionally, when events are partitioned by weapon type, statistically significant power laws persist (Fig.~\ref{fig:weapons}, Table~\ref{table:wsummary}) and show that any roughness in the scaling of the aggregate distributions (e.g., Fig.~\ref{fig:terrorism}a) is derived from the composition weapon-specific power laws with distinct scaling parameters and ranges. It also illustrates that there is something unique about explosives, which causes the shallow scaling of the lower tail for the injuries severity distribution.

There are many generative mechanisms in the literature for power laws, although many of them are unappealing for explaining the structure we find in global terrorism. The highly abstract model of competition between non-state actors and states, which we propose, analyze and extend via the mixtures model, is likely too simple to capture the fine structure of global terrorism. However, we hope that our model and the statistically significant empirical regularities which we show here will frame future efforts to understand global terrorism.

\vspace{4mm}
\noindent {\bf Acknowledgments.}~
The authors thank Cosma Shalizi, Raissa D'Souza and Cristopher Moore for helpful conversations. This work was supported in part  by the National Science Foundation under grants PHY-0200909 and ITR-0324845 (A.C.) and CCR-0313160 (M.Y.).


\end{document}